\documentclass[12pt,twoside]{IEEEtran+}
\usepackage{amsmath,amssymb}           
\def\BibTeX{{\rm B\kern-.05em{\sc i\kern-.025em b}\kern-.08em
    T\kern-.1667em\lower.7ex\hbox{E}\kern-.125emX}}
\def\={\triangleq}                           
\def\tr{\,{\rm tr}\,}
\newtheorem{thm}{Theorem}[section]
\newtheorem{defi}[thm]{Definition}
\newtheorem{rem}[thm]{Remark}
\newtheorem{cor}[thm]{Corollary}
\newtheorem{ex}[thm]{Example}
\newtheorem{lem}[thm]{Lemma}

\begin{document}

\title{Quantum Data Processing}
\author{Rudolf Ahlswede and Peter L\"ober\thanks{The authors are with
	Fakult\"at f\"ur Mathematik,
        Universit\"at Bielefeld, Postfach 100131, 33501 Bielefeld.}
        \thanks{E-mail: ahlswede@mathematik.uni-bielefeld.de, 
	                  loeber@mathematik.uni-bielefeld.de}}
\maketitle

\begin{abstract}
\noindent
We prove a data processing inequality for quantum communication channels,
which states that processing a received quantum state may never increase
the mutual information between input and output states.
\end{abstract}


\section{Introduction}
\noindent
Our starting point was the question whether Holevo's upper bound could be
viewed as a special case of a more general theorem, the quantum 
data processing inequality (that should hold). An inequality of this kind
states for an information transfer via (quantum) communication channels that
if a received state is ``processed'', this may never increase the
mutual information between input and output states.\\
We soon realized that a quantum data
processing inequality is a corollary of Uhlmann's Monotonicity Theorem (see
Sec.~\ref{UMT}) which is a generalization of a theorem of Lieb (\cite{Lie73}).
Like in the proof of the classical data processing inequality (which plays
around with conditional mutual information\footnote{cf.~\cite{CK81}, p.~55 or
\cite{CT91}, p.~32}), convexity properties play a central role. Here, we will
need such properties ``for operators'' (see Sec.~\ref{opMon}).\\


\section{Pick Functions}

\noindent
This section introduces the notion of Pick functions. We will see in the next
section that Pick functions are ``operator monotone''
(cf.~Cor.~\ref{PickFctsAreOpMon} for a rigorous formulation), a fact that is
very useful because often it's quite easy to decide whether a function
is a Pick function or not. The theory we introduce here is developed in great
detail (and with complete proofs) in \cite{Do70}.\\

\begin{defi}
$H^\pm \= \{x+iy\in{\mathbb C} | y\gtrless 0 \}$ denote the two half spaces,
respectively.\\
$P \= \{ \varphi\colon H^+ \to H^+ \ {\rm analytic}\ \}$ denotes the set of
\it Pick functions.\rm\\
\end{defi}

\begin{rem}
$P$ is a convex cone, and if $f,g\in P$ then $g\circ f\in P$, too.\\
\end{rem}

\begin{ex}
The function $\varphi(z) \= z^\mu$ with $0<\mu\le 1$ is in $P$.
A function 
$\psi(z) \= \alpha z+\beta+\sum_{i=1}^m \frac{\gamma_i}{\delta_i-z}$ with 
$\alpha,\gamma_i > 0$ and $\beta,\delta_i \in{\mathbb R}$ is in $P$, too.\\
\end{ex}

\noindent
The next theorem shows that the latter examples give essentially 
(i.e.~up to limits) all Pick functions:\\

\begin{thm}
Every Pick function $\varphi\in P$ has a (unique) representation
\begin{equation}\label{pickint}
\varphi(z) \= \alpha z+\beta+\int(\frac{1}{x-z}-\frac{x}{x^2+1})d\mu(x)
\end{equation}
with $\alpha\ge 0$, $\beta\in{\mathbb R}$ and $\mu$ a positive Borel measure
on ${\mathbb R}$ for which $\int(x^2+1)^{-1}d\mu(x) < \infty$.\\
\end{thm}

%

\noindent
This theorem is from \cite{Do70}, p.~20 ff. Its proof transforms the 
Pick function with an appropriate M\"obius transformation (and its inverse)
to a function which maps the unit disc into itself and which has a positive
real part. This real part is a positive harmonic function.\\

\begin{defi}
For an open interval $(a,b)\subseteq{\mathbb R}$ let ${\mathbb C}(a,b) \=
H^+ \cup H^- \cup (a,b)$ and
$$ P(a,b) \= \{ \varphi\colon{\mathbb C}(a,b)\to{\mathbb C} :
                \varphi|_{H^+}\in P \land 
                \varphi(\overline\cdot)=\overline{\varphi(\cdot)} \} \ .$$\\
\end{defi}

\begin{rem}
$P(a,b)$ is a convex cone. Moreover, $\varphi\in P(a,b)$ implies that 
$\varphi|_{(a,b)}$ is a monotonically increasing real function.
(As $\phi$ maps $H^+$ into $H^+$ and $H^-$ into $H^-$ it has to be real on the
real axis. Let now $\varphi=u+iv$ and $z=x+iy$. By definition
$\frac{d}{dy}v(x)\ge 0$, and the Cauchy-Riemann differential equations imply
that $\frac{d}{dx}u(x)\ge 0$, too.)\\
\end{rem}

\begin{ex}
The function $\varphi(z) \= z^\mu$ with $0<\mu\le 1$ is in $P(0,\infty)$.
A function 
$\psi(z) \= \alpha z+\beta+\sum_{i=1}^m \frac{\gamma_i}{\delta_i-z}$ with 
$\alpha,\gamma_i > 0$, $\beta\in{\mathbb R}$ and
$\delta_i\in{\mathbb R}\backslash(a,b)$ is in $P(a,b)$, too.\\
\end{ex}

\begin{rem}
Let $\varphi\in P$ a Pick function and $\mu$ its corresponding Borel measure
(cf.~(\ref{pickint})). Let $(a,b)\in{\mathbb R}$ an interval. Then:
$$ \varphi\in P(a,b) \qquad \Leftrightarrow \qquad \mu((a,b))=0 \ .$$\\
\end{rem}

\noindent
This remark is again from \cite{Do70} (p.~26), like the next example (p.~27):\\

\begin{ex}
$\sqrt{z} = \frac{1}{\sqrt{2}} + 
   \int_{-\infty}^0 (\frac{1}{x-z}-\frac{x}{x^2+1})\frac{\sqrt{x}}{\pi} dx$\\
\end{ex}


\section{Operator Monotonicity}\label{opMon}
\noindent
The first part of this section introduces the notion of operator monotonicity.
The results are taken from \cite{Do70}, pp.~67ff. We show that Pick
functions are operator monotone.\\
In the sequel we consider operator convexity properties like it was done in
\cite{HP82}, pp.~230ff. It will be important for the next sections that root 
functions are operator concave (cf.~Example~\ref{RootsAreOperatorConcave}).
Here, all Hilbert spaces (usually denoted by ${\cal H}$, etc.) are supposed to
be finite dimensional.\\

\begin{defi}
Let ${\cal H}$ a finite dimensional (complex) Hilbert space.
\begin{itemize}
\item ${\cal L}({\cal H})$ denotes the algebra of linear operators on ${\cal H}$. 
\item ${\cal L}({\cal H})^{s.a.}$ denotes the real vector space of self-adjoint
operators on ${\cal H}$. 
\item An operator $A\in{\cal L}({\cal H})$ is called
\it positive \rm if    $<v|Av>\,\ge 0\ \forall\,v\in{\cal H}$.
\item ${\cal L}({\cal H})^+$ denotes the convex cone of positive operators on
${\cal H}$.
(It holds ${\cal L}({\cal H})^+ \subset {\cal L}({\cal H})^{s.a.}$.)
\item For operators $A,B\in{\cal L}({\cal H})$ we write $A\le
B$ if $B-A$ is positive.\\
\end{itemize}
\end{defi}

\begin{defi}
Let ${\cal H}$ a finite dimensional (complex) Hilbert space,
$I\subseteq{\mathbb R}$ an interval, $f:I\to{\mathbb R}$ a function, and
$A\in{\cal L}({\cal H})^{s.a.}$ with all its eigenvalues in $I$.
\begin{itemize}
\item For $A=\sum_{i=1}^n a_i|a_i><a_i|$, where $(|a_i>)_{1\le i\le n}$ denotes
an ON basis of ${\cal H}$, we define $f(A)\=\sum_{i=1}^n f(a_i)|a_i><a_i|$.
(Clearly, this is independent of the chosen basis.) In matrix notation we have 
$$ A = \left(\begin{array}{rcl}a_1&&0\\&\ddots&\\0&&a_n\end{array} \right)
       \overset{f}{\mapsto} f(A)\ =
\left(\begin{array}{rcl}f(a_1)&&0\\&\ddots&\\0&&f(a_n)\end{array} \right) .$$
\item $f$ is called \it operator monotone of order $n$ \rm (``$f\in P_n(I)$'')
if $$ B\le C \quad \Longrightarrow \quad f(B)\le f(C) \qquad
\forall\,B,C\in{\cal L}({\cal H})^{s.a.}\ .$$
\item $f$ is called \it operator monotone \rm if
$f\in P_n(I)\ \forall\,n\in{\mathbb N}$.\\  
\end{itemize} 
\end{defi}

\begin{lem}
\begin{itemize}
\item $P_n(I)$ is a convex cone.
\item If $f\in P_n(I)$, $g\in P_n(J)$ with $im(f)\subseteq J$ then $g\circ f\in P_n(I)$.
\item $P_n(I)$ is closed (in the topology of pointwise convergence).
\item $P_{n+1}(I)\subseteq P_n(I)$.
\item For $\alpha>0$, $\beta\in{\mathbb R}$ it holds that $x\mapsto \alpha x+\beta$ is operator monotone.
\item $x \mapsto -\frac{1}{x}$ is operator monotone on $(0,\infty)$.\\
\end{itemize}
\end{lem}

\begin{proof}
All assertions but the last one are obvious.\\
So, for $A\in{\cal L}({\cal H})^+$ strictly positive ($0$ is no eigenvalue)
and $v,w\in{\cal H}$ it holds that 
\begin{eqnarray}
|<v|w>|^2 =
|<A^{-\frac{1}{2}}v|A^{\frac{1}{2}}w>|^2   &\le&
<A^{-\frac{1}{2}}v|A^{-\frac{1}{2}}v><A^{\frac{1}{2}}w|A^{\frac{1}{2}}w>\cr
&=& <v|A^{-1}v><w|Aw> \ , \nonumber
\end{eqnarray}
with equality (e.g.) for $w=A^{-1}v$. (This is
Cauchy-Schwarz-Buniakowski Inequality!) Therefore: 
$$ <v|A^{-1}v>\ =\,\max_{w\not=0} \frac{|<v|w>|^2}{<w|Aw>} \ ,$$ 
and this immediately implies that for $B,C\in{\cal L}({\cal H})^{s.a.}$: 
$$ B\le C \quad\Leftrightarrow\quad B^{-1}\ge C^{-1}  
\quad\Leftrightarrow\quad  -B^{-1}\le -C^{-1} \ .$$
\end{proof}

\begin{cor}\label{PickFctsAreOpMon}
Let $\varphi\in P(a,b)$. Then, $\varphi|_{(a,b)}$ is operator monotone.\\
\end{cor}


\begin{ex}
$f(x) \= x^\mu$ with $0\le\mu\le 1$ is operator monotone.\\
This is not (!) true if $\mu>1$. For (e.g.) $\mu=2$:
$$   \left(\begin{array}{cc}2 &2 \\2 &2 \end{array}\right)
  +  \left(\begin{array}{rr}1 &-1\\-1&1 \end{array}\right)
  =  \left(\begin{array}{cc}3 & 1\\1 &3 \end{array}\right)
 \le \left(\begin{array}{cc}3.1&0\\0&3.1\end{array}\right) \qquad {\rm but}$$
$$   \left(\begin{array}{cc}3 & 1\\1 &3 \end{array}\right)^2
  =  \left(\begin{array}{cc}10& 6\\6 &10\end{array}\right)\qquad{\rm and}\qquad
     \left(\begin{array}{cc}3.1&0\\0&3.1\end{array}\right)^2
  =  \left(\begin{array}{cc}9.61&0\\0&9.61\end{array}\right)\ .$$\\
\end{ex}

\begin{thm}
Let $f\ge 0$ a continuous and operator monotone function on $[0,\infty)$, and
let
$x\in{\cal L}({\cal H})^+$ and $a\in{\cal L}({\cal H})$ with $||a||_{op}\le 1$.
Then:
\begin{equation}\label{concavity}
f(a^*xa) \ge a^*f(x)a \ .
\end{equation}\\
\end{thm}

\begin{proof}
Let $b\=({\bf 1}-aa^*)^{\frac{1}{2}}$ and $c\=({\bf 1}-a^*a)^{\frac{1}{2}}$,
and define operators
$$ X \= \left(\begin{array}{cc}x &0 \\0 &0\end{array}\right), \qquad
   U \= \left(\begin{array}{cc}a&b\\c&-a^*\end{array}\right)$$
on ${\cal H} \oplus {\cal H}$. Given now $\varepsilon>0$ let $\lambda$ large
enough such that:
$$ Y \= \left(\begin{array}{cc}a^*xa+\varepsilon{\bf 1}&0 \\0 &\lambda{\bf 1}
              \end{array}\right)
 \ge \left(\begin{array}{cc}a^*xa&a^*xb\\bxa&bxb\end{array}\right) = U^*XU\ .$$
$$ \Longrightarrow \qquad f(Y) \ge F(U^*XU) = U^*f(X)U 
 \ge U^* \left(\begin{array}{cc}f(x)&0\\0&0\end{array}\right) U
 =  \left(\begin{array}{cc}a^*f(x)a&a^*f(x)b\\bf(x)a&bf(x)b\end{array}\right)$$
$$ \Longrightarrow \qquad a^*f(x)a \le f(a^*xa+\varepsilon{\bf 1})
         \underset{\varepsilon\to 0}{\longrightarrow} f(a^*xa) $$
\end{proof}

\begin{rem}
We call functions $f$ that fulfill (\ref{concavity}) \it operator concave. \rm
Indeed, a non-negative continuous function on $[0,\infty)$ that is operator
concave is also operator monotone (cf.~\cite{HP82}, p.~232). Furthermore,
operator concave functions $f$ fulfill \it Jensen's Inequality:\rm\\
$\forall\,x_1,\ldots,x_n\in{\cal L}({\cal H})^+$ and 
$a_1,\ldots,a_n\in{\cal L}({\cal H})$ with $\sum_i a_i^*a_i \le {\bf 1}$ 
it holds: $f(\sum_ia_i^*x_ia_i) \ge \sum_ia_i^*f(x_i)a_i$.\\
\end{rem}

\begin{proof}
$$X \= \left(\begin{array}{ccc}x_1&&0\\&\ddots&\\0&&x_n\end{array}\right),
\ A \=\left(\begin{array}{cc}a_1&\\\vdots&0\\a_n&\end{array}\right)
\ \Rightarrow\  A^*XA = \left(\begin{array}{cccc}
\sum_ia_i^*x_ia_i&0&\cdots&0\\0&&&\\\vdots&&0&\\0&&&\end{array}\right)$$
Therefore:\\
$$\left(\begin{array}{cccc}
f(\sum_ia_i^*x_ia_i)&0&\cdots&0\\0&&&\\\vdots&&0&\\0&&&\end{array}\right)
=f(A^*XA) \underset{(\ref{concavity})}{\ge} A^*f(X)A =\left(\begin{array}{cccc}
\sum_ia_i^*f(x_i)a_i&0&\cdots&0\\0&&&\\\vdots&&0&\\0&&&\end{array}\right) $$
\end{proof}

\begin{cor}\label{MonotonicityImpliesConcavity}
Let $f\ge 0$ a continuous and operator monotone function on $[0,\infty)$, and
let $x\in{\cal L}({\cal H})^+$ and $a\colon{\cal H'}\to{\cal H}$ a (lin.)
contraction ($||a||_{op}\le 1$). It holds: $f(a^*xa) \ge a^*f(x)a.$\\
\end{cor}

\begin{ex}\label{RootsAreOperatorConcave}
For $0\le\mu\le 1$ we have $(a^*xa)^\mu \ge a^*x^\mu a$ (with notation from
Cor.~\ref{MonotonicityImpliesConcavity}).\\
\end{ex}


\section{Finite Quantum Systems and Physical Maps}
\noindent
In this section we introduce the notion of finite quantum systems and
the maps of such systems that are considered to be in accordance with
quantum physics' laws.\\

\begin{defi}
A \it finite quantum system \rm ${\cal A}$ is a finite dimensional \it
$C^*$-algebra, \rm i.e. a self-adjoint subalgebra (with identity) of some 
${\cal L}({\cal H})$, where $\dim({\cal H})<\infty$.\\
A \it physical state \rm (on this system) is an element $A\in {\cal A}^+ \= 
{\cal L}({\cal H})^+ \cap {\cal A}$ for which $\tr(A)=1$.\\
\end{defi}

\begin{rem}
Let ${\cal A}={\cal L}({\cal H})$ and $A$ a physical state on ${\cal A}$.
Then $A=\left(\begin{array}{rcl}a_1&&0\\&\ddots&\\0&&a_n\end{array}\right)$
for an ONB of $A$-eigenvectors (by abuse of notation), and $A$ may be
interpreted as a probability distribution on these basis vectors.\\
\end{rem}

\begin{defi}
A ${\mathbb C}$-linear map $\alpha_*\colon{\cal A}\to{\cal B}$ is
\it positive \rm if $\alpha_*({\cal A}^+) \subseteq{\cal B}^+$. It is 
called \it completely positive \rm if all maps of the form
\begin{eqnarray*}
\alpha_*\otimes\mathbf 1\colon
{\cal A}\otimes{\cal C}&\to&{\cal B}\otimes{\cal C}\\
\sum_i a_i\otimes c_i &\mapsto& \alpha_*(a_i)\otimes c_i\\
\end{eqnarray*}
are positive. A \it physical map \rm of finite quantum systems 
${\cal A}, {\cal B}$ is a trace preserving and completely positive 
${\mathbb C}$-linear map $\alpha_*\colon{\cal A}\to{\cal B}$.\\
\end{defi}

\noindent
Fortunately, the completely positive maps have a nice representation by
Stinespring's Theorem (see \cite{Da76}, p.~137 or \cite{St55} for a proof):\\

\begin{thm}
Let $\alpha_*\colon{\cal A}\to{\cal L}({\cal H})$ a completely positive map
of finite quantum systems. Then
$$ \alpha_*(A) = V^*\rho(A)V \qquad (\forall\,A\in{\cal A}) $$
for some representation $\rho$ of ${\cal A}$ on a finite dimensional Hilbert
space ${\cal K}$ and a linear map $V\colon{\cal H}\to{\cal K}$. Here, the
representation $\rho$ is an (algebra-)homomorphism with $\rho(A^*)=\rho(A)^*$
for all $A\in{\cal A}$, and $\rho(\mathbf 1)=\mathbf 1$.\\
\end{thm}

\begin{rem}
A \it physical map \rm $\alpha_*\colon{\cal A}\to{\cal B}$ maps physical states
on ${\cal A}$ on physical states on ${\cal B}$. There are 3 equivalent ways to
describe a physical map of finite (!) quantum systems:
\begin{eqnarray}
(1)\qquad \alpha_*\colon{\cal A}&\to&{\cal B}\nonumber\\
                            A &\mapsto&\alpha_*(A)\nonumber\\
(2)\qquad \alpha\colon{\cal A}^*&\to&{\cal B}^*\nonumber\\
\tr_{\cal A}(A\ \cdot\,) &\mapsto&\tr_{\cal B}(\alpha_*(A)\ \cdot\,)\nonumber\\
(3)\qquad \alpha^*\colon{\cal B}&\to&{\cal A}\nonumber\\
&&\hspace{-2cm}\ldots\ {\rm such\ that}\ \alpha(\lambda)=\lambda\circ\alpha^*.
\nonumber
\end{eqnarray}
A ${\mathbb C}$-linear map $\alpha_*\colon{\cal A}\to{\cal B}$ is a physical
map if $\alpha^*$ is completely positive and unity preserving.
This implies that $\beta\=\alpha^*$ is a \it Schwarz map, \rm i.e.
$\beta(x^*x) \ge \beta(x)^*\beta(x)\ \forall\,x\in{\cal B}$. (This can be
deduced from Stinespring's Theorem.)\\
If $A\in{\cal A}^+$ is a physical state we call $\tr(A\ \cdot\,)\in{\cal A}^*$
a \it physical state, \rm too.\\
\end{rem}


\section{Uhlmann's Monotonicity Theorem}\label{UMT}
\noindent
Uhlmann's Monotonicity Theorem is our key tool to prove the quantum version of
a data processing inequality. Its proof (the proof of the following lemma) uses
the operator concavity of the root functions
(see Ex.~\ref{RootsAreOperatorConcave}).
For this section we closely followed \cite{OP93}, p.~18~ff.\\

\begin{lem}\label{UhlLem}
Let $\beta=\alpha^*\colon{\cal A}_1\to{\cal A}_2$ a physical map of finite
quantum systems, $S_1, T_1\in{\cal A}_1^+$ and $S_2, T_2\in{\cal A}_2^+$ with
the $T_i$ invertible ($i=1,2$). If for all $a\in{\cal A}_1^+$ it holds
$$ \tr(S_2\beta(a)) \le \tr(S_1a) \quad {\rm and} \quad
   \tr(T_2\beta(a)) \le \tr(T_1a),$$
then
$$\tr(\beta(x)^*S_2^t\beta(x)T_2^{1-t}) \le \tr(x^*S_1^txT_1^{1-t})
   \qquad \forall\ 0\le t\le 1, x\in{\cal A}_1\ .$$\\
\end{lem}

\begin{ex}
$\quad\tr(S_2^tT_2^{1-t}) \le \tr(S_1^tT_1^{1-t}).$\\
\end{ex}

\begin{proof}
We remark that ${\cal A}_1$ and ${\cal A}_2$ are Hilbert spaces with
inner product $<a,b>\ =\tr(a^*b)$.\\
Define a linear map $V\colon{\cal A}_1\to{\cal A}_2$ by $aT_1^{\frac{1}{2}}
\mapsto \beta(a)T_2^{\frac{1}{2}}$ for all $a\in{\cal A}_1$.\\
$V$ is a contraction, in fact:
$$ ||\beta(a)T_2^{\frac{1}{2}}||^2 = \tr(T_2\beta(a)^*\beta(a))
  \le \tr(T_2\beta(a^*a)) \le \tr(T_1a^*a) = ||aT_1^{\frac{1}{2}}||^2 .$$
For $i=1,2$ define $\Delta_i\in{\cal L}({\cal A}_i)^+$ by
$\Delta_i(aT_i^{\frac{1}{2}}) \= S_iaT_i^{-\frac{1}{2}}$ for all 
$a\in{\cal A}_i$.\\
(It is positive because $<aT_i^{\frac{1}{2}}|\Delta_i|aT_i^{\frac{1}{2}}>\
=\ <aT_i^{\frac{1}{2}}|S_iaT_i^{-\frac{1}{2}}>\
= \tr(T_i^{\frac{1}{2}}a^*S_iaT_i^{-\frac{1}{2}}) = \tr(a^*S_ia) \ge 0$.)\\
It holds $\Delta_i^t(aT_i^{\frac{1}{2}}) = S_i^taT_i^{\frac{1}{2}-t}$
($t\ge 0$), and $V^*\Delta_2V \le \Delta_1$:
$$<aT_1^{\frac{1}{2}}|V^*\Delta_2V|aT_1^{\frac{1}{2}}>\
=\ <\beta(a)T_2^{\frac{1}{2}}|\Delta_2|\beta(a)T_2^{\frac{1}{2}}>\
=\ <\beta(a)T_2^{\frac{1}{2}}|S_2\beta(a)T_2^{-\frac{1}{2}}>$$
$$=\tr(\beta(\alpha)^*S_2\beta(a)) \le \tr(S_2\beta(\alpha\alpha^*))
  \le \tr(S_1\alpha\alpha^*) 
   =\ <aT_1^{\frac{1}{2}}|\Delta_1|aT_1^{\frac{1}{2}}>.$$
So\footnote{We use Ex.~\ref{RootsAreOperatorConcave} (like we promised
above).}, $V^*\Delta_2^tV \le (V^*\Delta_2V)^t \le \Delta_1^t$, and
\begin{eqnarray}
\tr(T_2^{\frac{1}{2}}\beta(x)^*S_2^t\beta(x)T_2^{\frac{1}{2}-t})
 &=&\ <xT_1^{\frac{1}{2}}|V^*\Delta_2^tV|xT_1^{\frac{1}{2}}>\nonumber\\
 &\le&\ <xT_1^{\frac{1}{2}}|\Delta_1^t|xT_1^{\frac{1}{2}}>\
 \ \ = \ \ \tr(T_1^{\frac{1}{2}}x^*S_1^txT_1^{\frac{1}{2}-t})\ .\nonumber
\end{eqnarray}
\end{proof}

\begin{defi}
Given a physical state $A\in{\cal A}^+$ we denote in the sequel its \it 
support, \rm i.e.~the projector on the space spanned by the eigenvectors 
corresponding to non-zero eigenvalues, by $supp\ A$.\\
Two physical states $A,B\in{\cal A}^+$ have \it divergence \rm
$$ D(A||B) \= 
\begin{cases}
\tr(A(\log A-\log B)), &{\rm if}\ supp\ A \le supp\ B\\
\infty,                & otherwise.
\end{cases}$$
Equivalently, physical states $\omega=\tr(A\ \cdot\,),\varphi=\tr(B\ \cdot\,)
\in {\cal A}^*$ have \it divergence \rm $D(\omega||\varphi)\=D(A||B)$.
Sometimes, divergence is called \it relative entropy, \rm too.\\
\end{defi}

\noindent We are ready for Uhlmann's Monotonicity Theorem (\cite{Uhl77}):\\

\begin{thm}\label{UhlThm}
Let $\beta=\alpha^*\colon{\cal A}_1\to{\cal A}_2$ a physical map of finite
quantum systems and $\omega,\varphi\in {\cal A}_2^*$ physical states. Then:
$$ D(\omega\circ\beta||\varphi\circ\beta) \le D(\omega||\varphi)\ .$$\\
\end{thm}

\begin{proof}
Let $\omega=\tr(S_2\ \cdot\,)$, $\varphi=\tr(T_2\ \cdot\,)$,
$\omega\circ\beta=\tr(S_1\ \cdot\,)$ and $\varphi\circ\beta=\tr(T_1\ \cdot\,)$,
where w.l.o.g. $supp\ T_2={\bf 1}$.\\
As $\tr(S_2\beta(\cdot)) = \omega\circ\beta=\tr(S_1\ \cdot\,)$ and
$\tr(T_2\beta(\cdot)) = \varphi\circ\beta=\tr(T_1\ \cdot\,)$
Lemma \ref{UhlLem} implies that:
$$ \tr(S_2^\mu T_2^{1-\mu}) \le \tr(S_1^\mu T_1^{1-\mu}) 
   \qquad  (\forall\,0\le\mu\le 1). $$
Consequently, $\frac{1-\tr(S_2^\mu T_2^{1-\mu})}{1-\mu} \ge
\frac{1-\tr(S_1^\mu T_1^{1-\mu})}{1-\mu}$, and by the limit $\mu\to 1$:
$$ D(\omega||\varphi) = \tr(S_2^\mu T_2^{1-\mu})'|_{\mu\to 1}
 \ge \tr(S_1^\mu T_1^{1-\mu})'|_{\mu\to 1}
  =  D(\omega\circ\beta||\varphi\circ\beta)\ .$$
\end{proof}


\section{The Quantum Data Processing Inequality}
\noindent
In this final section we derive the Quantum Data Processing Inequality from
Uhlmann's Monotonicity Theorem. We start with a modified formulation of the
latter:\\

\begin{cor}
Let $\alpha_*\colon{\cal A}\to{\cal B}$ a physical map of finite quantum 
systems and $A_1, A_2\in{\cal A}^+$ physical states. Then:
$$ D(\alpha_*(A_1)||\alpha_*(A_2)) \le D(A_1||A_2).$$\\
\end{cor}

\begin{proof}
This is only a question of notation:
\begin{eqnarray}
D(A_1||A_2) &=& D(\tr(A_1\ \cdot\,)||\tr(A_2\ \cdot\,)),\ {\rm and}\nonumber\\
D(\alpha_*(A_1)||\alpha_*(A_2)) &=&
D(\tr(\alpha_*(A_1)\ \cdot\,)||\tr(\alpha_*(A_2)\ \cdot\,)).\nonumber
\end{eqnarray}
To reduce the claim to Theorem \ref{UhlThm} notice that
$\tr(\alpha_*(A_i)\ \cdot\,) = \alpha(\tr(A_i\ \cdot\,))
= \tr(A_i\alpha^*(\cdot))$.\\
\end{proof}

\begin{defi}
Let $W_*\colon{\cal A}\to{\cal B}$ a quantum channel, i.e.~a physical map
of finite quantum systems. Let $A\in{\cal A}^+$ a physical state with
$A=\sum_a a|a><a|$ its spectral decomposition, and let $B\=W_*(A)$ the 
corresponding output (physical) state. Furthermore, let $(A,B)
\=\sum_a a|a><a|\otimes W_*(|a><a|)$
the joint state. The \it mutual information \rm is given by
$$ I(A;W_*) \= H(A)+H(B)-H(A,B) $$
where $H(X)\=-\tr(X\log X)$ denotes the Shannon-von Neumann entropy.\\
\end{defi}

\noindent Here is the Quantum Data Processing Inequality:\\

\begin{cor}
Let $W_*\colon{\cal A}_1\to{\cal A}_2$ and  $D_*\colon{\cal A}_2\to{\cal A}_3$
physical maps of finite quantum systems and $A\in{\cal A}_1^+$ a physical
state. Then:
$$ I(A;D_* \circ W_*) \le I(A;W_*) \ .$$\\
\end{cor}

\begin{proof}
Because of $ I(A;W_*) = D((A,B)||A\otimes B)$ this is just a consequence of
$$ D(({\bf 1}\otimes D_*)(A,B)||({\bf 1}\otimes D_*)(A\otimes B))
   \ \le\ D((A,B)||A\otimes B) \ .$$
\end{proof}

\noindent As special case we get \it Holevo's Upper Bound \rm (\cite{Hol73}):\\
 
\begin{cor}
Let $W_*\colon{\cal A}_1\to{\cal A}_2$ a quantum channel and
$D_*$ a measurement on its output space, i.e.~a physical map from ${\cal A}_2$
into some commutative $C^*$-algebra ${\cal A}_3$. Then 
$I(A;D_* \circ W_*) \le I(A;W_*)$.\\
\end{cor}

\noindent
We remind the reader that Holevo's result is from 1973 whereas Uhlmann's
Monotonicity Theorem is from 1977. Both use analytical considerations for
the proofs. There is also an ``elementary'' proof of Holevo's Upper Bound,
using only information theoretical considerations, in \cite{Win99}.\\


\section{Acknowledgments}
\noindent
We thank Andreas Winter for fruitful discussions about 
quantum information theory.\\


\end{document}